# Simulation-based Prediction of Black Hole X-Ray Spectra and Spectral Variability

Rongrong Liu[1], Chris Nagele[1], Julian H. Krolik[1], Brooks E. Kinch[2], and Jeremy D. Schnittman[3]
[1] Department of Physics and Astronomy, Johns Hopkins University, Baltimore, MD 21218, USA
[2] Department of Mechanical Engineering and Applied Mechanics, University of Pennsylvania, Philadelphia, PA 19104, USA
[3] Gravitational Astrophysics Lab, NASA Goddard Space Flight Center, Greenbelt, MD 20771, USA

## Abstract

Data derived from general relativistic magnetohydrodynamic simulations of accretion onto black holes can be used as input to a postprocessing scheme that predicts the radiated spectrum. Combining a relativistic Compton scattering radiation transfer solution in the corona with detailed local atmosphere solutions incorporating local ionization and thermal balance within the disk photosphere, it is possible to study both spectral formation and intrinsic spectral variability in the radiation from relativistic accretion disks. With this method, we find that radiatively efficient systems with black holes of $10M_{\odot}$ accreting at $\approx$0.01 in Eddington units produce spectra very similar to those observed in the hard states of X-ray binaries. The spectral shape above 10 keV is well described by a power law with an exponential cutoff. Intrinsic turbulent variations lead to order-unity changes in bolometric luminosity, variations in the logarithmic spectral slope $\sim$0.1, and factor of 2 alterations in the cutoff energy on timescales $\sim$50 $(M_{\rm BH}/10M_{\odot})$ ms. Within the corona, the range of gas temperature spans more than 1 order of magnitude. The wide distribution of temperatures is central to defining the spectrum: the logarithmic spectral slope is harder by $\sim$0.3 and the cutoff energy larger by a factor $\sim$10–30 than if the coronal temperature everywhere were its mass-weighted mean.

## 1. Introduction

"Coronal" X-ray emission is a nearly ubiquitous feature of accreting black holes. From Galactic black hole X-ray binaries in which the black hole mass is $\sim 10M_{\odot}$ to quasars with black hole masses $\sim 10^9 M_{\odot}$, power-law spectra running from a few keV up to a crudely exponential cutoff at $\sim$100 keV can frequently be seen. Interestingly, the power-law indices for the "hard" state in X-ray binaries and for active galactic nuclei (AGN) span almost the same limited range: if the spectrum is described in energy units, i.e., energy flux per unit photon energy $F_{\epsilon} \propto \epsilon^{-\alpha}$, the observed range for stellar-mass black holes is $0.4 < \alpha < 1.1$ (R. A. Remillard & J. E. McClintock 2006), and for AGN with black hole masses many orders of magnitude larger, it is $0.6 < \alpha < 1.5$ (A. C. Fabian et al. 2017; A. Tortosa et al. 2018). Variability is also a hallmark of this emission, both in luminosity (R. A. Remillard & J. E. McClintock 2006) and in spectral shape (M. Zhou et al. 2022; I. Pal et al. 2024; S. Wang et al. 2024).

Although the energy source for the hard X-ray luminosity of accreting black holes has long been thought to be dissipation of magnetic field rising buoyantly within the accretion flow (E. P. T. Liang & R. H. Price 1977; A. A. Galeev et al. 1979; F. Haardt & L. Maraschi 1991), attempts to understand this radiation have largely focused on phenomenological models in which single homogeneous regions containing electrons at a certain temperature Compton-scatter seed photons originally radiated by a nearby thermal region in the flow, and observational data are described in similarly phenomenological fashion (e.g., as in the XSPEC model `comptt` based on L. Titarchuk 1994). Very little attention has been given either to how the spectrum might be affected by the accretion flow's internal structure or to how its variability is driven by the flow's internal dynamics, other than through other phenomenological models (Y. E. Lyubarskii 1997; A. R. Ingram 2016; P. Uttley & J. Malzac 2025; S. G. D. Turner & C. S. Reynolds 2023; but see M. Moscibrodzka et al. (2007) for explicit calculation of spectral variability in 2D global simulations of optically thin accretion, S. C. Noble & J. H. Krolik (2009) for a discussion of the origin of coronal luminosity fluctuations in distributed MHD turbulence, and P. C. Fragile et al. (2023) for some data on spectral variations associated with MHD turbulence).

Here we adopt a different point of view—one that seeks to predict both the spectral shape of this emission and its variability properties directly from explicit dynamical calculations. It thereby automatically incorporates both spatial inhomogeneity and variability grounded in the underlying physics rather than suppositions. By this means, in this paper we make the first exploration of how dynamical processes within an accretion flow lead to spectral variations in hard X-ray radiation.

Our method begins with general relativistic magnetohydrodynamic (GRMHD) accretion simulations incorporating a zeroth-order approximation to Compton cooling in the corona and a measure of the local cooling rate in the disk body. We then postprocess 3D simulation snapshots of rest-mass density, 4-velocity, internal energy density, and heating rate. Electron temperature and photon spectral intensity in the corona are calculated with a relativistic Monte Carlo radiation transport code that enforces local thermal balance; these solutions are then made consistent with a large set of 1D radiation transfer solutions covering the entire disk body outside any thermodynamic equilibrium regions. The opacities and emissivities in the disk body are computed from local photionization and thermal balance, with the radiation field's outer boundary condition taken from the Monte Carlo solution giving the local coronal X-ray illumination. This part of the procedure computes both the "reprocessed" emission and the intrinsic emission of the disk self-consistently. The result is a prediction of the entire system's radiated spectrum as a function of both photon energy and direction.

An overview of this method is given in Section 2. Our results (plots of spectra and power-law fits) are presented in Section 3. Their significance is discussed in Section 4.

## 2. Methods

Our work is built upon the framework developed in B. E. Kinch et al. (2016, 2019), which uses snapshots of accreting stellar-mass black holes produced by 3D GRMHD simulations to predict the X-ray spectra seen by a distant observer (J. D. Schnittman & J. H. Krolik 2013; J. D. Schnittman et al. 2013). We extend the original framework to examine the time-dependent characteristics of black hole X-ray spectra by analyzing multiple snapshots from a single simulation, separated by a fixed interval. The procedure can be summarized as follows:

Using HARM3D (S. C. Noble et al. 2009), a GRMHD code designed to simulate the dynamics of accreting material around black holes in 3D space, we generate 3D maps of its rest-mass density, internal energy density, 4-velocity, and cooling rate.

The original HARM3D code (S. C. Noble et al. 2009) used a target temperature cooling function to control the disk thickness: when bound matter exceeded a certain target temperature $T_*$, it was cooled back to $T_*$, achieving a desired disk aspect ratio H/R. After evolving for $\gtrsim$10,000$M$, where $M$ is the unit of time for $G = c = 1$, accretion disks treated this way reached a state in which the disk within $r \approx 25M$ was in inflow equilibrium and therefore in a steady-state with respect to its mass profile. In physical units, of $M = (M_{\rm BH}/M_\odot) \times 4.9 \times 10^{-6}$ s, and distance can likewise be measured in units of $M = (M_{\rm BH}/M_\odot) \times 1.5 \times 10^5$ cm.

For the results presented here, we restarted a simulation with spin parameter $a/M = 0.9$ and H/R $= 0.06$ at $t = 10{,}000M$ (J. D. Schnittman et al. 2016). Within its Compton scattering photosphere, we continued to use the "target temperature" cooling function just described, but outside the photosphere this was replaced with the coronal cooling function described in B. E. Kinch et al. (2020). The photosphere is defined as the location where the Thomson optical depth reaches unity when integrated away from the polar axis along polar angle coordinate curves. Instead of forcing the structure to a desired H/R, the coronal cooling function approximates the local cooling effects of inverse Compton scattering. It is calculated and applied every time step (1 time step $\approx 1.5 \times 10^{-3}M$).

The principal approximation in this method is that the radiation energy density in each cell ($u_{\rm rad}$) is estimated assuming the photons in that cell have traveled there from the disk photosphere following straight lines in coordinate space and without alteration in their energy. As shown in B. E. Kinch et al. (2020), this procedure yields a radiation energy density that is a good approximation to an exact solution. Because summing over the large number of contributing rays is computationally expensive, whereas $u_{\rm rad}$ changes very little in a single fluid time step, we update $u_{\rm rad}$ only every 100 time steps (we compared this recalculation interval to shorter ones and found very little difference in the results).

Because the coronal cooling rate depends on the Compton opacity, the unit of gas density is no longer arbitrary and instead depends on $\dot{m}$, the mass accretion rate in Eddington units. For this reason, $\dot{m}$ must be specified within the simulation. For the results presented here, we chose $\dot{m} = 0.01$. However, because distance and time are defined in terms of the black hole mass $M$, full conversion from scale-free code units to physical units (cgs) can be done in the postprocessing stage after choosing a mass for the central black hole $M$. The conversion relationships for density $\rho$ and cooling function $\mathcal{L}$ are (J. D. Schnittman et al. 2013)

$$\rho_{\rm cgs} = \rho_{\rm code} \frac{4\pi c^2}{\kappa G M} \frac{\dot{m}/\eta}{\dot{M}_{\rm code}}, \quad (1)$$

$$\mathcal{L}_{\rm cgs} = \mathcal{L}_{\rm code} \frac{4\pi c^7}{\kappa G^2 M^2} \frac{\dot{m}/\eta}{\dot{M}_{\rm code}}. \quad (2)$$

Here $\kappa = 0.4\ {\rm cm^2\ g^{-1}}$ is the electron scattering opacity and $\eta$ is the radiative efficiency ($\eta = 0.1558$ in this simulation, chosen to be the binding energy of an ISCO orbit for $a/M = 0.9$). Note that the choice of $\eta$ determines the units of the problem but not the actual luminosity or efficiency because they are calculated by the simulation. For the results presented here, $M = 10M_\odot$. Future work will treat a wider range of black hole masses.

After the restart with the coronal cooling function, we ran the simulation for a further 6000$M$, allowing us to obtain seven snapshots separated in time by 1000$M$. These snapshots are the data on which our present work rests.

Each of these snapshots was then postprocessed with `Pandurata` in order to obtain the emergent spectrum (J. D. Schnittman & J. H. Krolik 2013; B. E. Kinch et al. 2016). First the location of the photosphere was determined for each ($r$, $\phi$) in the grid by integrating the Thomson opacity over polar angle from the nearest polar cutout until reaching unit optical depth. Then the cooling rate was integrated between the top and bottom photospheric surfaces for each ($r$, $\phi$) to determine the effective temperature at the photosphere. Next, photon packets with a blackbody spectrum corresponding to the local effective temperature were injected from the photosphere into the corona to initiate a Monte Carlo transfer solution in which photons travel along relativistic geodesics and undergo Compton scattering. Both angular and energy redistribution are evaluated including Klein–Nishina effects, a relativistic thermal energy distribution for the electrons, fully relativistic scattering kinematics, and the transformations between the coordinate frame and the fluid frame. When a Monte Carlo solution is completed, we group the cells into $3 \times 3 \times 3$ "sectors" in order to suppress statistical fluctuations in the number of scattering events in the individual cells. Comparing the cooling rate in each sector to its radiation energy density we find the electron temperature $T_e$ at which Compton cooling would account for the cooling rate. The temperature used for the previous Monte Carlo solution is then replaced by $T_e$, and a new Monte Carlo solution is run. We deem a sector to be converged when the cooling power in that sector from `Pandurata` matches the cooling power from `HARM3D` to within 1%. Global temperature convergence is reached when either $\geqslant$99% of the sectors have converged or $\geqslant$95% of sectors have converged and the total coronal cooling power matches to within 1%.

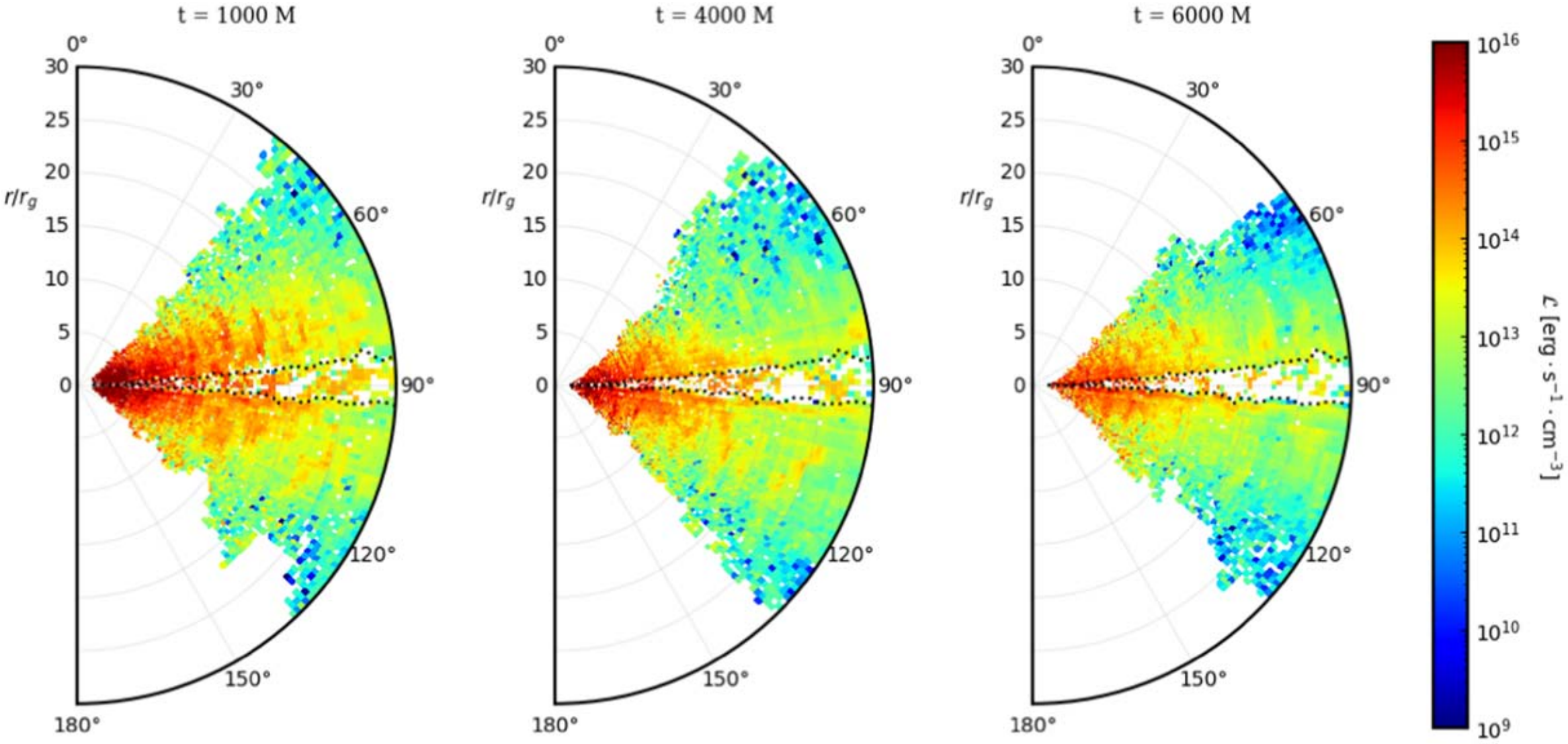

**Figure 1.** Vertical slices of the cooling rate in the simulation at times $t = 1000M$, $4000M$, and $6000M$. The dotted black line shows the disk photosphere; the white region surrounding the polar axis is the region where no scatterings occurred, so the Monte Carlo solution showed no Compton cooling.

After the coronal temperature map has converged, the photon flux on the surface of the disk is used as a boundary condition for the solution of a 1D atmosphere solution at each cell on the disk photosphere. In this solution, the radiation transfer problem, including atomic opacities and emissivities as well as Compton opacity, is solved by the Feautrier method in conjunction with a local solution of the thermal balance and ionization equilibrium equations for all elements with great enough abundance to be relevant (B. E. Kinch et al. 2016, 2019). In the thermal balance, the cooling rate recorded from the HARM3D simulation is treated as a local heating rate because it is set so as to capture the local dissipation. When the combined transfer, thermal balance, and ionization equilibrium solution has converged, the surface spectrum for each of the photospheric cells is reevaluated and a new Monte Carlo solution for the corona is obtained. This new solution also incorporates an energy-dependent albedo for the disk surface constructed using the opacities within the disk. After several cycles between coronal and disk solutions, the entire system converges to equilibrium, i.e., two successive cycles yield spectra that agree to within 1% at all photon energies >1 keV.

The very final step is to collect the photon packets headed toward viewers on a solid angle grid of 40 polar angles and 39 azimuthal angles in order to evaluate the spectrum as seen at infinity, including all relevant Doppler shifts. The spectra we show here are integrated over solid angle. Due to the projected size of the accretion disk and corona, the radiated flux declines by a factor ∼2 from a view along the polar axis to one that runs just outside the disk photosphere, but there is essentially no change in spectral shape as a function of azimuthal angle.

Two further remarks about the limitations of our method remain to be made. The first has to do with the low-energy end of our predicted spectra. These photons are produced primarily by the disk surface at large radii, where the effective temperature is comparatively low. However, the underlying simulation falls out of inflow equilibrium at large radii, making both the cooling rate and the structure of the disk in those locations questionable. For this reason, we do not regard the predicted spectra below a certain energy to be entirely reliable. The second has to do with the opposite end of the temperature range. In portions of the corona, particularly those near the polar axis and far from the disk surface, the temperature is likely to be very high but ill defined because there are very few photon scattering events in the Monte Carlo solution: these regions have very low density gas density, and few photons from the disk photosphere pass through them. As a result, they are not represented in the outgoing spectrum. In addition, even where there are Monte Carlo scattering events, if the temperature rises to $kT_e \gtrsim m_e c^2$, copious pair production should occur, a process not currently included in our method's physics repertory. Pair production generically diminishes the temperature by converting heat into pair rest mass, likely keeping the temperature in the ∼100–500 keV range (as discussed in, e.g., J. H. Krolik 1999). At present, however, there are very few observations of spectra in the 100 keV–1 MeV band, limiting comparison of our predictions to data.

## 3. Results

### 3.1. Thermal Properties

Before discussing how much the spectrum varies over time, it is worthwhile to present the fundamental data on which the spectral properties depend: maps of cooling rate and equilibrium temperature in the corona.

The local cooling rate plays a critical role in determining both the equilibrium temperature and, of course, the bolometric emissivity of the plasma. Its variation in terms of location and time determines the character of spectral variation. Figure 1 illustrates the cooling rate's intermittency, both in space and in time. In the corona, there is a 4 orders of magnitude contrast from greatest cooling rate per volume to least. It is generally largest just outside the disk body and at small radii. Overall, it diminishes with both increasing angle from the midplane and radius, but this diminution is quite irregular: it can vary by 1 order of magnitude or more on length scales as small as $\sim M$ and by order unity on timescales as short as $\sim 1000M$. The cooling rate within $\sim 40°$ of the polar axis is effectively zero because there were no scattering events in the Monte Carlo transfer solution within that cone. Inside the disk body, the cooling rate is even more intermittent and irregular. At any given time, the cooling rate is negligible in most of the volume but very large in a few places.

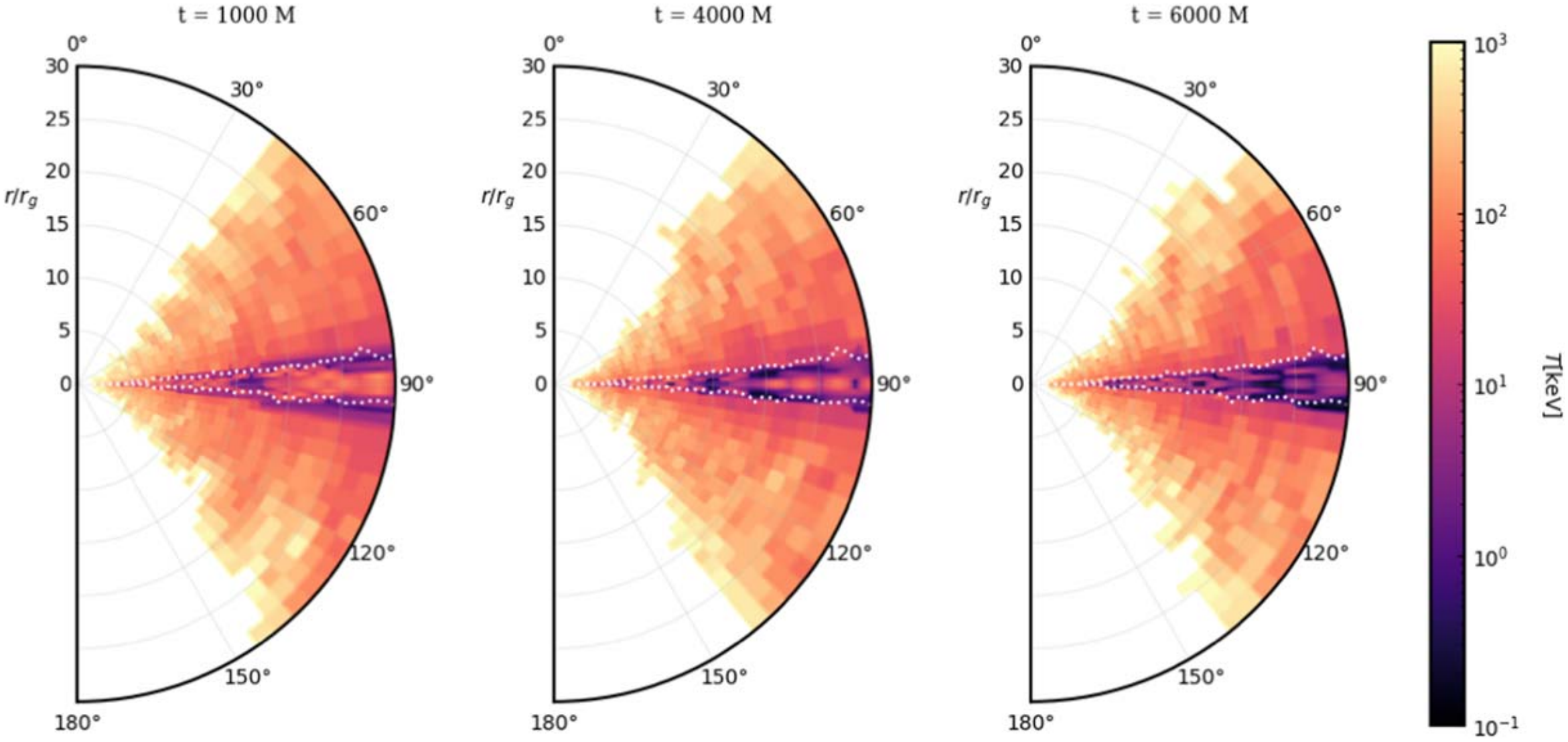

**Figure 2.** Equilibrium temperature in the same slices shown in Figure 1. The dotted white line marks the photosphere.

Figure 2 shows the equilibrium temperature on a single poloidal slice obtained from converged Monte Carlo transfer solutions at the same three times shown in Figure 1. Strikingly, in all cases, the temperature depends only weakly on radius, instead varying primarily with polar angle. The temperature also spans a very large range, from $\sim$1 MeV at the hottest places with a defined temperature to $\sim$10–30 keV just outside the disk photosphere to $\sim$100 eV at the coldest spots inside the disk body. No temperature can be defined within the cone where no scattering events took place because both the Compton scattering power and the radiation energy density there are undefined. Although the temperature's dependence on polar angle is crudely monotonic in the corona, it is more complicated inside the disk body due to the complex interaction of highly intermittent turbulent dissipation with the inward diffusion of photons boosted to high energy in the corona.

### 3.2. Emitted Spectra

Our predicted spectra as seen at infinity and integrated over solid angle are displayed in Figure 3; the corresponding bolometric luminosity for each time is shown in Figure 4. From Figure 4, we see that during this relatively short span of time ($6000M = 0.3(M/10M_{\odot})$ s) the luminosity varied by a factor $\sim$2.7, but in this brief time segment there is little variability power on still shorter timescales.

From Figure 3, we see that the magnitude of variation is strongly dependent on photon energy, with the lower energies changing much more weakly than the higher energies, especially those near the peaks of the spectra ($\sim$100–500 keV). The contrast at 300 keV between the most and least luminous times is a factor of 5, while at 10 keV the drop from brightest to faintest is only a factor $\sim$3, and at 1 keV, it is even smaller, a factor $\sim$1.7.

The small variation at the lowest energies is partially physical and partially an artifact. Because the low-energy part of the spectrum is made primarily at larger radii in the disk, all dynamical processes are slower, leading to limited changes in output over the $6000M$ time span under consideration. In addition, the specific character of the spectrum below $\sim$10 keV is not reliably portrayed here because over the $\approx$16,000$M$ of the entire simulation, those same larger radii (i.e., $r \gtrsim 30$–$35M$) had not yet arrived at inflow equilibrium, and photons created in the disk body at those radii contribute significantly to that energy band. For this reason, in this paper we will focus on the higher-energy portion of the spectrum, photons $\gtrsim$10 keV.

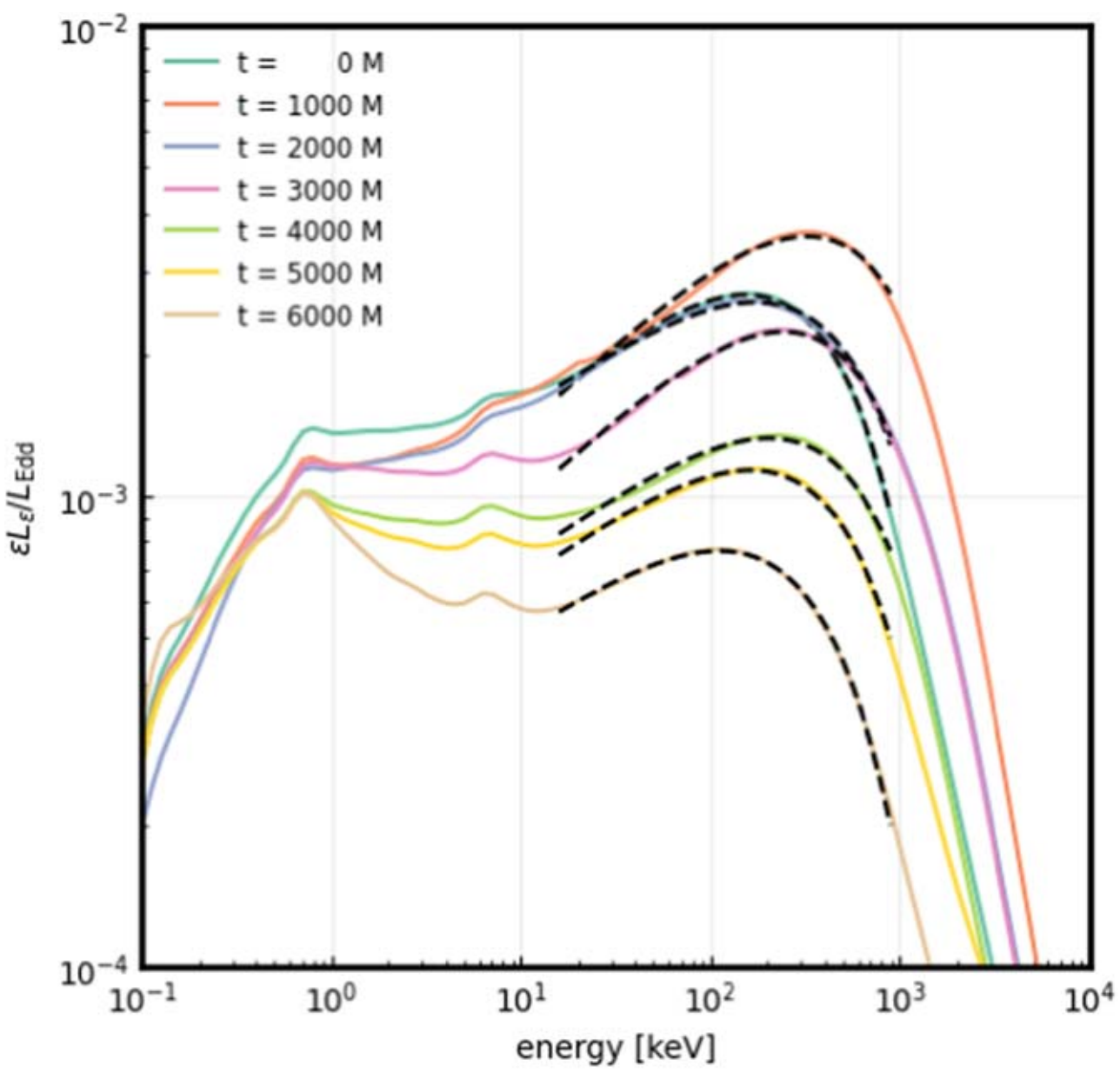

**Figure 3.** Spectra at seven times spaced $1000M$ apart. Dashed lines show the best-fit exponential cutoff power law for each.

To describe the high-energy spectrum, we fit it to a simple analytic form, an exponential cutoff power law of the form $\epsilon L_\epsilon/L_{\rm edd} = A\epsilon^b \exp(-\epsilon/\epsilon_{\rm cut-off})$. With this parameterization, $b \equiv d\ln F/d\ln\epsilon$, the flux power-law index $\alpha \equiv d\ln F_\epsilon/d\ln\epsilon = b - 1$. Observations are often fit to a form $\propto \epsilon^{-\Gamma}$ to describe the number of photons per unit energy; this index $\Gamma = 2 - b$. Minimizing the rms deviation between this form and our computed spectra from 16 keV to 1 MeV yields the results shown in Table 1. It is clear from Figure 3 that this form provides a very good fit to the computed spectrum up to energies $\simeq 2\epsilon_{\rm cutoff}$ in every case.

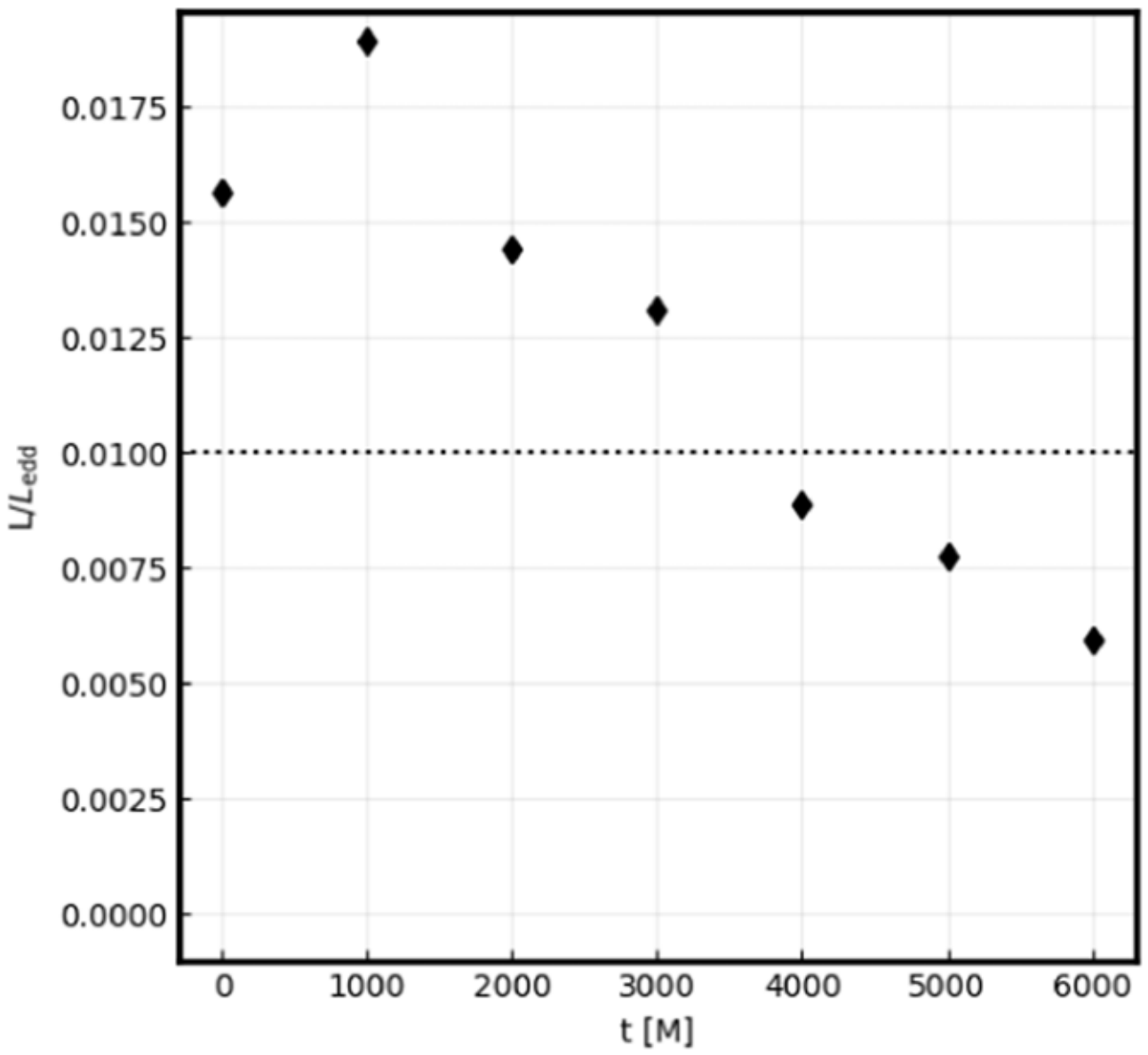

**Figure 4.** Bolometric luminosity as a function of time.

**Table 1**
Fitting Coefficients for Spectral Lightcurves at Different Times

| Time ($M$) | Fitting Coefficients | | |
|---|---|---|---|
| | $A$ | $b$ | $\epsilon_{\rm cutoff}$ (keV) |
| 0 | $6.03 \times 10^{-5}$ | 0.347 | 440 |
| 1000 | $4.28 \times 10^{-5}$ | 0.379 | 852 |
| 2000 | $12 \times 10^{-5}$ | 0.277 | 623 |
| 3000 | $3.12 \times 10^{-5}$ | 0.375 | 646 |
| 4000 | $5.23 \times 10^{-5}$ | 0.288 | 704 |
| 5000 | $4.55 \times 10^{-5}$ | 0.293 | 553 |
| 6000 | $4.24 \times 10^{-5}$ | 0.272 | 408 |

Even across these rather short timescales—$1000M \approx 50(M/10M_\odot)$ ms—the exponent $b$ and the cutoff energy $\epsilon_{\rm cutoff}$ both change. The logarithmic slope $b$ varies modestly, ranging from 0.22 to 0.36. On the other hand, the cutoff energy varies by more than a factor of 2, from $\simeq$400 to $\simeq$900 keV. In this small sample, there is a weak correlation between luminosity and $\epsilon_{\rm cutoff}$, but there is little correlation between luminosity and $b$.

We close this section with a few remarks about the two distinct emission-line features evident at lower energies. One, a broad bump from $\sim$4 to 8 keV, is associated with the Fe K$\alpha$ fluorescence line. In this case, it is primarily radiated by highly ionized Fe-, He-, and H-like ions. The other, at $\sim$0.7–0.9 keV, may be Ne X emission lines, transitions within the Fe L shell in highly ionized (but not stripped) Fe, or a blend of numerous other lines. With longer runs, it will be possible to evaluate their equivalent widths more quantitatively.

## 4. Discussion

### 4.1. Physical Determination of the Instantaneous Spectrum

To understand these variations it is necessary to consider what underlies the parameters of these fits. The spectrum generated by the entire accretion flow is jointly produced by processes creating, absorbing, and reprocessing photons in the disk body and Compton scattering of those photons in the corona. In the case treated here, the majority of the total heating rate is in the corona, so our discussion will focus on Comptonization.

Treatments of spectral formation by Compton scattering often assume, for understandable reasons, that the region in which it takes place is homogeneous in temperature. When the Thomson optical depth $\tau_T \lesssim 1$, the Zel'dovich argument then yields a cutoff power law whose index depends separately on $\tau_T$ and the electron temperature $T_e$ (G. B. Rybicki & A. P. Lightman 1985; J. H. Krolik 1999) and is exponentially cut off on the $kT_e$ scale. Where the Kompane'ets equation is more appropriate, a thermal cutoff power law also emerges, with the index dependent on the Compton-$y$ parameter, $y \equiv (kT_e/m_e c^2)\max(\tau_T, \tau_T^2)$ (S. L. Shapiro et al. 1976).

However, our far more physical treatment demonstrates that the optical depth from any given point in the corona to the outside can vary from 0 to 1, while there are significant mass fractions in the corona at temperatures ranging from $\sim$7 to $\sim$70 keV (see Figure 5), and the range of temperatures contributing significantly to the luminosity runs from $\sim$25 keV to $\sim$2 MeV, peaking at different times anywhere from $\sim$100 to $\sim$500 keV (see Figure 5 or B. E. Kinch et al. 2021). As remarked earlier, in the hottest regions, our omission of pair physics likely leads to an overestimate of $T_e$, so the true temperature range may be somewhat narrower on the upper end. These broad temperature distributions strongly suggest that coronae are fundamentally inhomogeneous in terms of temperature. Not only do they contain wide ranges of temperature; as illustrated in Figure 5, the distribution of power with temperature is offset substantially from the distribution of mass with temperature.

To gauge how well or poorly a single temperature can perform as an indicator of the spectrum produced, we compare the spectrum for the $t = 1000M$ snapshot that would be produced by a fully structured inhomogeneous corona with that produced by a corona with the same spatial distribution of electrons but a single temperature. Perhaps the two most physically justified weighted means are those with respect to mass and with respect to heating power. In this case, the mass-weighted mean temperature is $\simeq$26 keV, while the power-weighted mean is $\simeq$500 keV. The factor of $\simeq$20 between these two definitions of the mean temperature itself signals the importance of the wide span of temperatures present in the corona.

The resulting spectra, along with the spectrum derived from the actual temperature distribution, are displayed in Figure 6. As it shows, when the temperature throughout the corona is the mass-weighted mean, the spectrum at energies $\lesssim$10 keV is close to the actual spectrum but diverges sharply at higher energies, with the mass-weighted mean temperature model consistently underpredicting the luminosity. A corona with the mass-weighted mean temperature creates a spectrum with roughly constant $dL/d\ln\epsilon$ for $\epsilon \lesssim 60$ keV and drops sharply above that energy, whereas a corona with the actual temperature spread produces a spectrum in which $dL/d\ln\epsilon$ rises steadily from $\approx$8 keV to a peak at $\approx$400 keV. On the other hand, when the temperature throughout is the power-weighted mean (500 keV), the spectrum is dramatically harder across the entire range of photon energies. Not surprisingly, the lower uniform temperature also produces a bolometric luminosity well below the actual value, while the higher uniform temperature produces a bolometric luminosity far above the actual one. It is therefore clear that neither of these two averaging schemes can correctly describe the photon output.

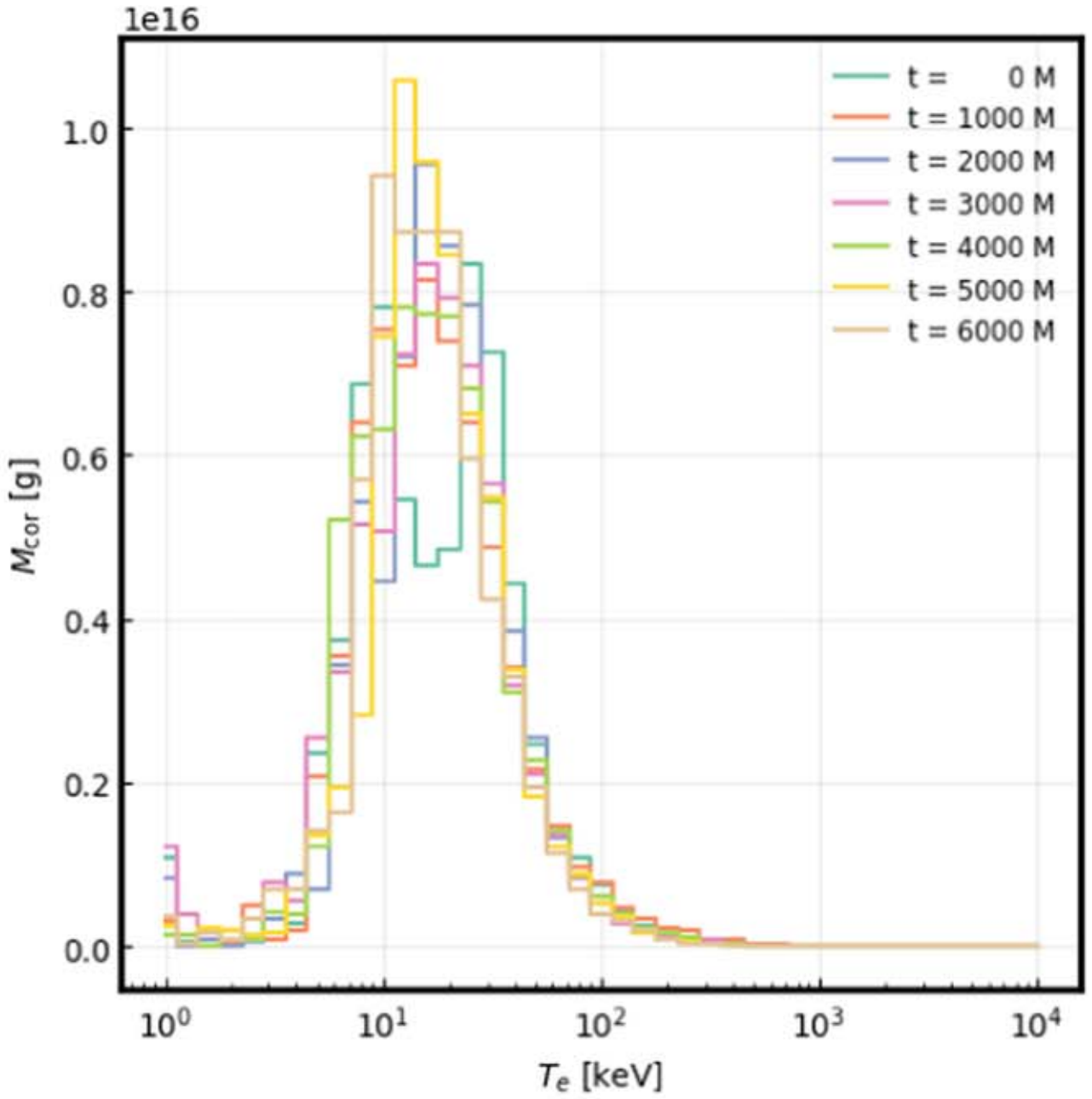

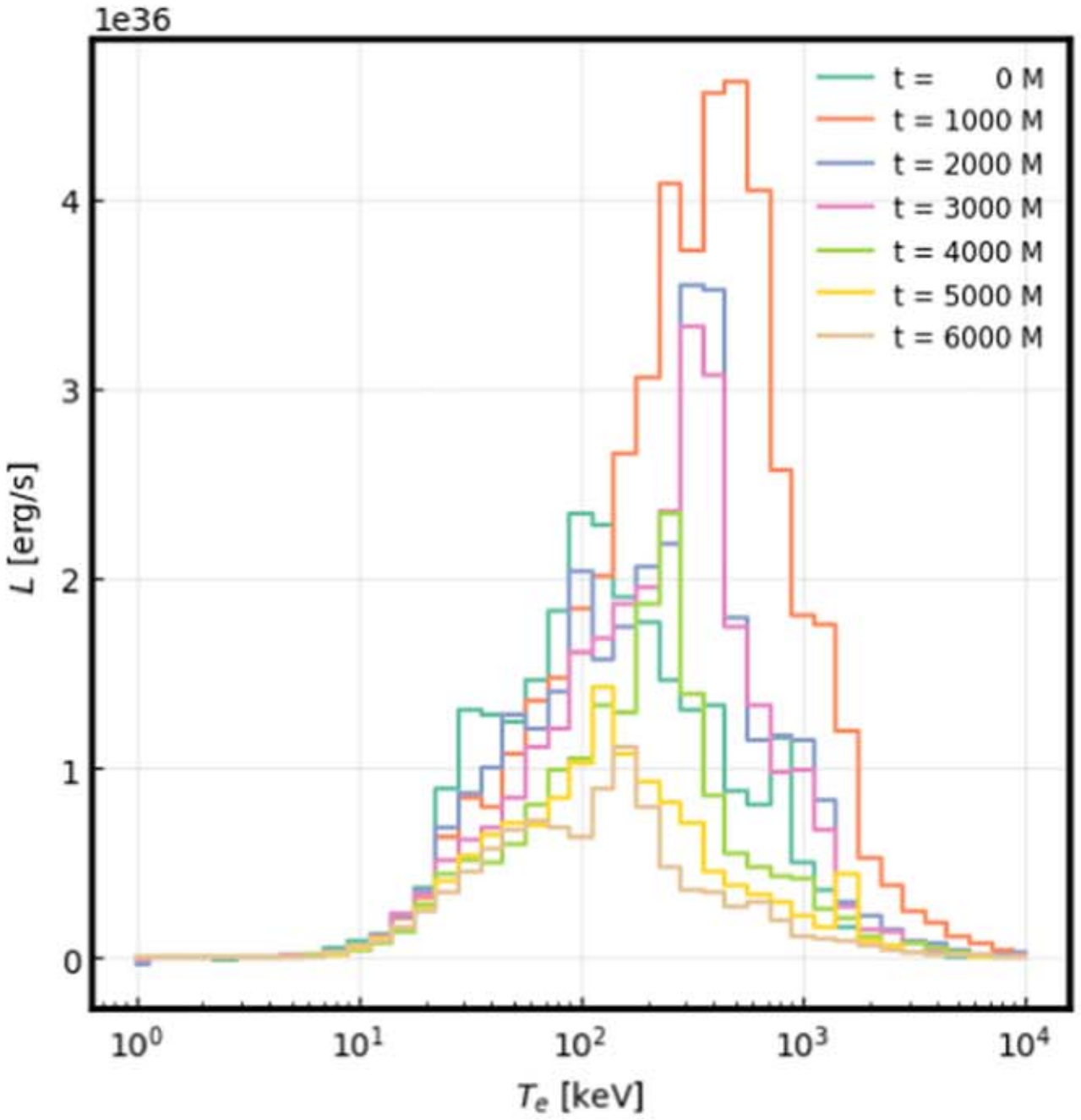

**Figure 5.** Distributions of mass with temperature (top) and radiated energy with temperature (bottom). Units are per bin, i.e., $\partial L/\partial \log T_e$ and $\partial M \partial \log T_e$. For both distributions, the data are at $t = 1000M$ and the bins have width $\Delta \log T_e = 0.1$.

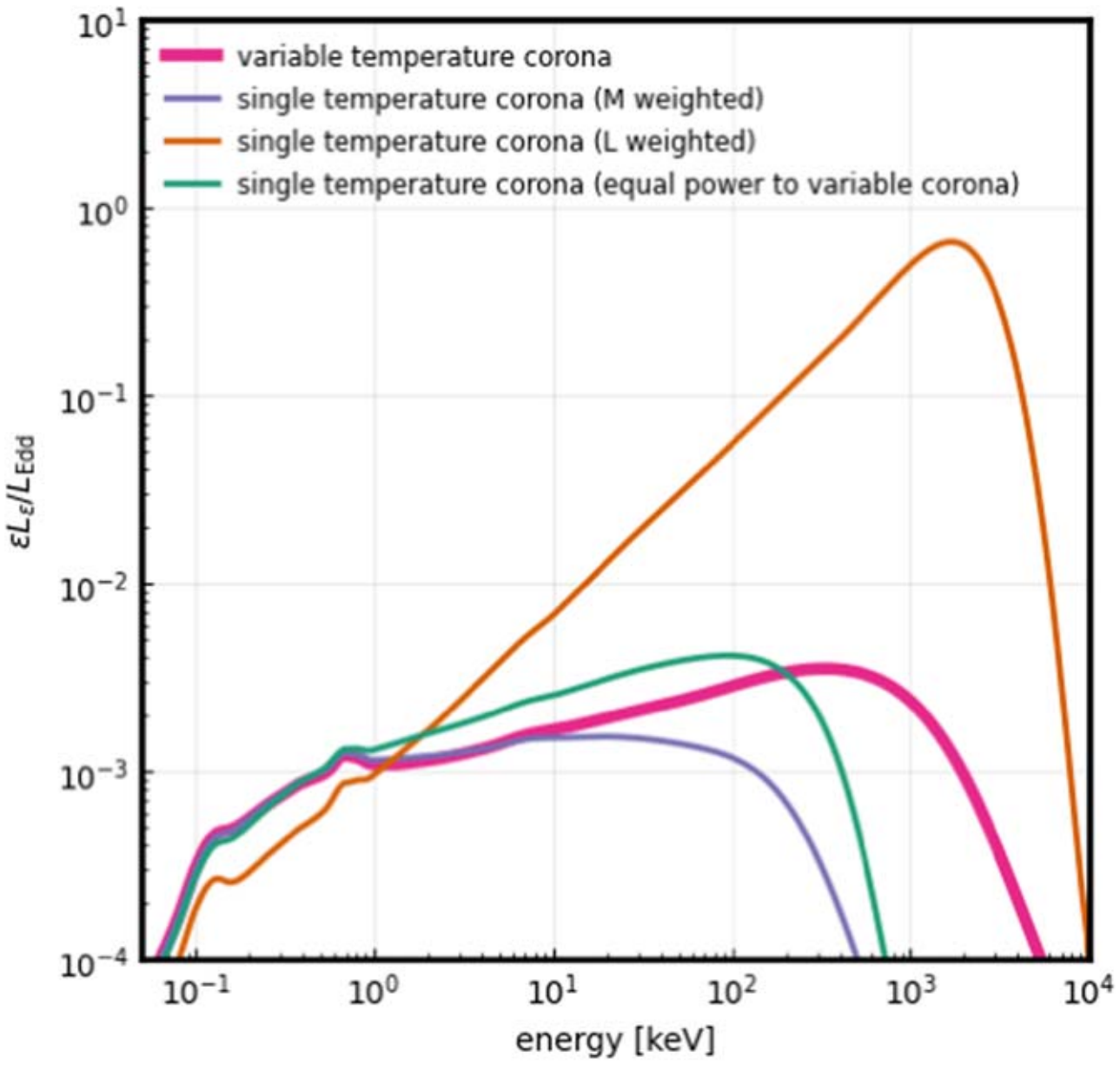

**Figure 6.** Emergent spectrum. Actual spectrum for $1000M$ as calculated by our method (magenta thick curve) and the spectrum calculated from a corona with the same density distribution but uniform temperature set to the mass-weighted mean of 28 keV (purple), the heating-weighted mean of 500 keV (orange), or the fixed-luminosity temperature of 54 keV (green).

One might also ask the question, "Is it possible to find any single temperature for the corona at which it radiates the correct luminosity?" This can, in fact, be done. In the present instance, again using the $t = 1000M$ data, the luminosity-matching temperature is $\simeq$54 keV. Somewhat surprisingly, not only does it yield the correct total luminosity (as it was constrained to do), but it also produces very nearly the correct logarithmic slope (again see Figure 6). However, its high-energy cutoff is a factor $\simeq$6 too small, $\simeq$150 keV rather than $\simeq$940 keV. Thus, it is possible to find a single temperature such that both the emitted luminosity and the spectrum's power-law index are correct but at the price of a badly wrong cutoff energy. Moreover, the averaging method that might define this temperature is obscure.

To understand better why the spectrum due to a distribution of temperatures cannot be reproduced by any single-temperature model, one may think of spectral formation in terms of the total energy amplification for a photon emerging from the corona. This total amplification is the product of all the individual scattering events' amplifications. When the amplification is the same everywhere (as it would be in a homogeneous-temperature corona in which the photon energies are always $\ll m_e c^2$), the total amplification is simply $A^k$, where $A$ is the individual scattering amplification and $k$ is the number of scatters. By contrast, when the temperature is inhomogeneous, the total amplification, even for paths with the same number of scatters, can be quite different for different paths, and this statement remains true even when the temperature map is constructed, as in our method, to yield the correct luminosity. Thus, the emergent spectrum depends strongly on the specific map of temperature, and if there is a fairly broad range of temperatures, there is no single-temperature model that can adequately represent the emergent spectrum.

This last point leads to a final remark. We have seen that when the corona has a single temperature, its high-energy cutoff energy is several times the temperature. This is possible because photons arrive at these energies after multiple scatters, so photon energies of several $kT_e$ can be the result of multiple interactions with electrons, each of which adds an energy $\lesssim kT_e$. This result extends to the more physical situation on which we focus, in which the corona spans a wide range of temperatures. Examination of the distribution of cooling power with temperature shown in Figure 5 reveals that the high-energy cutoff $\epsilon_{\rm cutoff}$ is found at an energy several times the temperature at which the distribution of cooling with temperature peaks. With a spread in temperatures, the ratio between $\epsilon_{\rm cutoff}$ and the temperature of the peak increases (decreases) when the distribution $\partial L/\partial \log T_e$ declines more gradually on the high-temperature (low-temperature) side of the peak.

### 4.2. Comparison to Observed Spectra and Parameter Inference

It is noteworthy that the spectral slopes found from first principles in a radiatively efficient general relativistic MHD simulation in which $\dot{m} = 0.01$ and $M_{\rm BH} = 10 M_\odot$ fall squarely within the range observed in X-ray binaries (which is also,more or less the range observed in AGN, where $M_{\rm BH}$ is much larger). Our predicted cutoff energies, $\epsilon_{\rm cutoff} \approx 400$–900 keV, are a factor $\sim$2–4 larger than measured in, for example, GRO J1655-40 (N. Shaposhnikov et al. 2007); as noted earlier, the absence of pair production from our calculation likely overestimates $kT_e$ when it is $\gtrsim m_e c^2$. Although hard states are often described by phenomenological models in which the spectrum is formed in a radiatively inefficient and optically thin region occupying the inner radii of the accretion flow (A. A. Esin et al. 1997; J. Ferreira et al. 2006; C. Done et al. 2007), we find that it is also possible to produce a hard-state X-ray spectrum without any radiatively inefficient zone because a sizable fraction of the total dissipation takes place in the corona above an optically thick disk.

This situation may not be universal, but it is reasonably common. For spin 0.9 and $\dot{m} = 0.01$, the coronal dissipation fraction in the underlying GRMHD simulation on which we base these results is $\simeq$85%; for the same spin and $\dot{m} = 0.1$, or for spin 0.5 and $\dot{m} = 0.01$, the fraction is only somewhat smaller, $\simeq$70% (B. E. Kinch et al. 2021). Observations support the suggestion that a radiatively efficient disk often accompanies the coronal region (R. C. Reis et al. 2010; J. M. Miller et al. 2015; X. Q. Ren et al. 2022).

The importance of the broad span of temperatures in the corona to both the spectral slope and the high-energy cutoff means that it is misleading to use either quantity to infer a single value for the coronal temperature. This point is seen most dramatically in the interpretation of $\epsilon_{\rm cutoff}$; it is a characteristic energy for the spectral shape, but its relation to the corona's temperature is more indirect. For example, the spectrum emitted at $t = 1000M$ in our simulation yielded a cutoff energy $\approx$900 keV even though the mass-weighted mean temperature of the corona was only 26 keV. In fact, the cutoff energy is also $\sim 2\times$ the temperature at which the power distribution peaks; this becomes possible both because of the multiple scatterings already mentioned and because the distribution of cooling with temperature extends beyond the temperature at the distribution's peak.

### 4.3. Interpretation of Time-integrated Spectra

As already mentioned, sizable changes in both luminosity and spectral shape can occur on timescales as short as $\sim 50(M/10M_\odot)$ ms; such variations therefore probe frequencies as high as $\sim 10(M/10M_\odot)^{-1}$ Hz. This frequency range is at the high end of what is currently feasible to probe in X-ray binaries: NICER observations of 0.5–10 keV X-rays from Cyg X-1 while in the "hard" state have probed up to 100 Hz (maximum bin duration 10 ms; O. König et al. 2024), but most lightcurves (obtained on less sensitive instruments observing fainter sources) are constrained to sampling at considerably longer timescales. It follows that the spectra they measure, particularly at energies $\gtrsim$10 keV, are time integrals summing over many instantaneous spectra like those portrayed in Figure 3. Such sums tend to be dominated by the brightest epochs because they contribute disproportionately to the total.

### 4.4. Dependence of Variability Amplitude on Photon Energy

Spectral variability at the frequencies explored here ($\sim 2-10(M/10M_\odot)^{-1}$ Hz) is strongest at energies $\gtrsim$10 keV, and this fact can be attributed to these variations being driven by fluctuations in the relatively small amount of high-temperature gas. This prediction is consistent with observations in which the fluctuation power in the 5–10 keV band becomes noticeably larger than in the 1.5–5 keV band at frequencies $\gtrsim$5 Hz (T. Kawamura et al. 2022).

### 4.5. Timescale for Order-unity Luminosity Variations

As demonstrated here, intrinsic fluctuations in coronal heating are capable of altering its bolometric luminosity by factors of several on timescales $\sim 1000M$. Because our underlying simulation depends on $\dot{m}$ but not $M_{\rm BH}$, our findings about variability can be tested against observations of hard X-ray luminosity in systems spanning a wide range of mass and luminosity. For example, the mass of the black hole in Cyg X-1 is $21 \pm 2 M_\odot$ (J. C. A. Miller-Jones et al. 2021), so our timescale $1000M = 0.11$ s; consistent to within a factor $\sim$2, I. M. McHardy et al. (2004) found a "knee" in the hard-state fluctuation power spectrum at $\approx$23 Hz, above which the fractional fluctuation amplitude per log frequency dropped below $\sim$15%. In the Seyfert galaxy NGC 4051, I. M. McHardy et al. (2004) found a similar break in the fluctuation power spectrum at $\approx$0.8 mHz. If the characteristic frequency of variation reflects a pure linear scaling with black hole mass, this measurement implies a mass $\sim 6 \times 10^5 M_\odot$, close to the "reverberation mapping" estimate of $5 \times 10^5 M_\odot$ (M. M. Fausnaugh et al. 2017; an estimate with its own potential systematic errors: see J. H. Krolik 2001, S. Casura et al. 2024). Another Seyfert galaxy, NGC 5548, has exhibited order-unity X-ray fluctuations on a timescale $\sim 5 \times 10^5$ s (J. S. Kaastra et al. 2014); if identified with a time $1000M$, this translates to $M_{\rm BH} \sim 1 \times 10^8 M_\odot$. This, too, is similar to the most recent reverberation-estimated mass, $\simeq 1.4 \times 10^8 M_\odot$ (K.-X. Lu et al. 2022), but it should be acknowledged that in the many reverberation campaigns mounted on this source, the product $r(\Delta v)^2$ has varied over a factor of 3, and different authors' "virial factors" have varied over a similar range.

## 5. Conclusions

Building on the work of B. E. Kinch et al. (2016, 2019) and especially B. E. Kinch et al. (2021), we have studied in detail how the properties of a relativistic accretion flow exhibiting strong MHD turbulence can lead directly to spectra closely resembling those of black hole X-ray binaries in the hard state. When the accretion rate in Eddington units has a value in the range often seen in hard states, the bulk of the dissipation takes place in the corona. Its temperature ranges from $\sim$10 keV just outside the disk's photosphere to $\sim$1 MeV (neglecting pair production, which would reduce the maximum temperature).

Although most of the mass exists near the low end of the temperature range, the hotter gas is the site of a disproportionate amount of dissipation. For this reason, the power law produced by Comptonization is considerably hotter than what would be produced by a homogeneous corona whose temperature is the mass-weighted value, and its cutoff energy is much larger than would be predicted by the mass-weighted temperature.

Because of the intrinsic turbulent fluctuations, the power-law slope and the cutoff energy both vary on timescales $\sim 1000M \sim 50(M/10M_{\odot})$ ms. The range of variation in the slope is modest ($\sim \pm 0.15$). In our calculations, the range of variation in cutoff energy is much greater (a factor of $\sim 2$), but this range may be diminished when pair production is taken into account.

Perhaps most importantly, we have also demonstrated that the shape of the spectrum produced by coronal inverse Compton scattering cannot be understood in terms of homogeneous-temperature models in which the temperature represents a mean over the intrinsically wide distribution within the corona. Large internal temperature contrasts are intrinsic features of accretion disk coronae, and they produce spectra different from regions with uniform temperature.

## Acknowledgments

We thank Scott Noble, the author of HARM3D, for use of his code and for much help in running it.

This work was partially supported by NSF Grant AST-2009260 and NASA TCAN grant 80NSSC24K0100.

## ORCID iDs

Rongrong Liu https://orcid.org/0000-0003-0685-3525
Julian H. Krolik https://orcid.org/0000-0002-2995-7717
Brooks E. Kinch https://orcid.org/0000-0002-8676-425X
Jeremy D. Schnittman https://orcid.org/0000-0002-2942-8399

## References

Casura, S., Ilić, D., Targaczewski, J., Rakić, N., & Liske, J. 2024, MNRAS, 534, 182
Done, C., Gierliński, M., & Kubota, A. 2007, A&ARv, 15, 1
Esin, A. A., McClintock, J. E., & Narayan, R. 1997, ApJ, 489, 865
Fabian, A. C., Lohfink, A., Belmont, R., Malzac, J., & Coppi, P. 2017, MNRAS, 467, 2566
Fausnaugh, M. M., Grier, C. J., Bentz, M. C., et al. 2017, ApJ, 840, 97
Ferreira, J., Petrucci, P. O., Henri, G., Saugé, L., & Pelletier, G. 2006, A&A, 447, 813
Fragile, P. C., Anninos, P., Roth, N., & Mishra, B. 2023, ApJ, 959, 59
Galeev, A. A., Rosner, R., & Vaiana, G. S. 1979, ApJ, 229, 318
Haardt, F., & Maraschi, L. 1991, ApJL, 380, L51
Ingram, A. R. 2016, AN, 337, 385
Kaastra, J. S., Kriss, G. A., Cappi, M., et al. 2014, Sci, 345, 64
Kawamura, T., Axelsson, M., Done, C., & Takahashi, T. 2022, MNRAS, 511, 536
Kinch, B. E., Noble, S. C., Schnittman, J. D., & Krolik, J. H. 2020, ApJ, 904, 117
Kinch, B. E., Schnittman, J. D., Kallman, T. R., & Krolik, J. H. 2016, ApJ, 826, 52
Kinch, B. E., Schnittman, J. D., Kallman, T. R., & Krolik, J. H. 2019, ApJ, 873, 71
Kinch, B. E., Schnittman, J. D., Noble, S. C., Kallman, T. R., & Krolik, J. H. 2021, ApJ, 922, 270
König, O., Mastroserio, G., Dauser, T., et al. 2024, A&A, 687, A284
Krolik, J. H. 1999, Active Galactic Nuclei (Princeton, NJ: Princeton Univ. Press)
Krolik, J. H. 2001, ApJ, 551, 72
Liang, E. P. T., & Price, R. H. 1977, ApJ, 218, 247
Lu, K.-X., Bai, J.-M., Wang, J.-M., et al. 2022, ApJS, 263, 10
Lyubarskii, Y. E. 1997, MNRAS, 292, 679
McHardy, I. M., Papadakis, I. E., Uttley, P., Page, M. J., & Mason, K. O. 2004, MNRAS, 348, 783
Miller, J. M., Tomsick, J. A., Bachetti, M., et al. 2015, ApJL, 799, L6
Miller-Jones, J. C. A., Bahramian, A., Orosz, J. A., et al. 2021, Sci, 371, 1046
Moscibrodzka, M., Proga, D., Czerny, B., & Siemiginowska, A. 2007, A&A, 474, 1
Noble, S. C., & Krolik, J. H. 2009, ApJ, 703, 964
Noble, S. C., Krolik, J. H., & Hawley, J. F. 2009, ApJ, 692, 411
Pal, I., Anju, A., Sreehari, H., et al. 2024, ApJ, 976, 145
Reis, R. C., Fabian, A. C., & Miller, J. M. 2010, MNRAS, 402, 836
Remillard, R. A., & McClintock, J. E. 2006, ARA&A, 44, 49
Ren, X. Q., Wang, Y., Zhang, S. N., et al. 2022, ApJ, 932, 66
Rybicki, G. B., & Lightman, A. P. 1985, Radiative Processes in Astrophysics (New York: Wiley)
Schnittman, J. D., & Krolik, J. H. 2013, ApJ, 777, 11
Schnittman, J. D., Krolik, J. H., & Noble, S. C. 2013, ApJ, 769, 156
Schnittman, J. D., Krolik, J. H., & Noble, S. C. 2016, ApJ, 819, 48
Shapiro, S. L., Lightman, A. P., & Eardley, D. M. 1976, ApJ, 204, 187
Shaposhnikov, N., Swank, J., Shrader, C. R., et al. 2007, ApJ, 655, 434
Titarchuk, L. 1994, ApJ, 434, 570
Tortosa, A., Bianchi, S., Marinucci, A., Matt, G., & Petrucci, P. O. 2018, A&A, 614, A37
Turner, S. G. D., & Reynolds, C. S. 2023, MNRAS, 525, 2287
Uttley, P., & Malzac, J. 2025, MNRAS, 536, 3284
Wang, S., Brandt, W. N., Luo, B., et al. 2024, ApJ, 974, 2
Zhou, M., Grinberg, V., Bu, Q. C., et al. 2022, A&A, 666, A172